\documentclass[prl,twocolumn,showpacs,amsmath,amssymb]{revtex4}

\usepackage{bm}
\usepackage{dcolumn}
\usepackage[dvipdfm]{graphicx}

\usepackage{color}

\begin{document}

\preprint{preprint}

\title{
\textcolor{black}{Protection of the surface states in topological insulators:
Berry phase perspective}
}
\author{Ken-Ichiro Imura$^1$}
\author{Yositake Takane$^1$}
\affiliation{$^1$Department of Quantum Matter, AdSM, Hiroshima University, Higashi-Hiroshima 739-8530, Japan}

\date{\today}

\begin{abstract}
The metallic surface state of a topological insulator (TI) is not only topologically protected, 
\textcolor{black}{
but exhibits a remarkable property of inducing an effective vector potential}
on curved surfaces. 
For an electron in the surface state of
a spherical or a cylindrical TI (TI nanoparticle or nanowire)
a pseudo-magnetic monopole or a fictitious solenoid is effectively induced,
encoding the geometry of the system.
\textcolor{black}{
Here, by taking an example of a hyperbolic surface we demonstrate that
as a consequence of this property stemming from its active spin degree of freedom,
the surface state is by itself topologically protected.}
\end{abstract}


\maketitle

\begin{figure}[t]
\begin{tabular}{c}
\includegraphics[width=45mm]{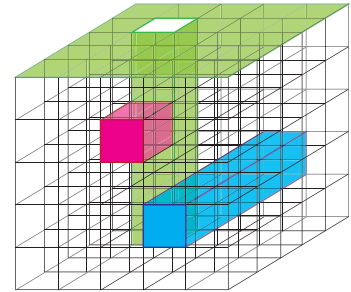}
\end{tabular}
\vspace{-2mm}
\caption{(Color online)
Which is the genuine surface?}
\label{JG}
\end{figure}

Neither being a metal nor an insulator, 
the topological insulator has now been recognized
as a basic form of solid 
that exhibits both gapped bulk and gapless surface states
\cite{FuKaneMele, MooreBalents, Roy}.
Such a classification is well-defined in the continuum limit, 
while the situation is less trivial in the case of lattice models 
often employed as a concrete implementation of topological insulators,
there remaining a question, ``where actually is the surface?''
A lattice model is sparse, 
and in a somewhat extreme point of view,
existing only on sites and links,
so that
its surface is not restricted to the macroscopic boundary of the system,
but could be also chosen
{\it e.g.},
such that it is partly extended to a rectangular-prism-shaped region (RPSR)
penetrating into the bulk
as depicted Fig. \ref{JG}.
Or one can also think of an atomic scale closed surface 
isolated in the bulk
\cite{closed}.
However, in reality the protected surface state appears 
only on its macroscopic surface, 
exhibiting no symptom of penetrating into the bulk even in the case of sparse lattice systems.

Why is the surface state 
\textcolor{black}{thus} {\it noninvasive} into the bulk? 
What prevents it from penetrating into the sparsely filled interior of the lattice models? 
In this Communication we demonstrate that
the existence of a Berry phase $\pi$,
or a spin connection associated with what is often called spin-to-surface locking
\cite{Ran_PRB, Vishwanath, Mirlin, Bardarson, disloc, aniso},
plays a central role in this issue.
Though existence of a protected surface state is a defining property 
of the topological insulator,
topological protection does not exclude the possibility of finite-size gap opening.
As we have demonstrated previously
\cite{aniso, spherical},
Dirac electrons on the surface of a topological insulator
\textcolor{black}{encodes information on
the geometry of the sample
in the form of spin connection 
that appears in the effective surface Dirac Hamiltonian.}
On a cylindrical surface
a fictitious solenoid threading the cylinder is effectively induced
\cite{aniso},
while in the case of a spherical system,
an effective magnetic monopole
\cite{spherical, Shen}
is induced,
determining the gapped electronic spectrum on the surface.

A Dirac electron on the surface of a topological insulator, 
especially, its spin state is susceptible of two types of constraints, 
and ``locked'' both in the momentum and real spaces.
Spin-to-momentum locking is a direct consequence of the strong spin-orbit coupling 
in this system. 
Here, we focus on
another phenomenon that manifests on a curved surface,
often represented by the term, ``spin-to-surface locking''.
Through the bulk-boundary correspondence,
the entangled nature of the spin and the configuration spaces
encoded in the bulk Hamiltonian
is transcribed to the surface Dirac equation.
The helical surface state thus inherits
a geometrical constraint imposed on its spin state,
and an electron in this state
is susceptible of a specific type of Berry phase, or the spin connection,
inducing an effective monopole or a flux tube.
In the somewhat special case of cylindrical geometry, 
the constraint on spin manifests as spin-to-surface locking,
{\it i.e.},
the spin of the surface state is constrained 
onto the tangential plane of the curved surface
\cite{Ran_PRB, Vishwanath, Mirlin, Bardarson, disloc, aniso}.
Appearance of the spin connection in the surface Dirac equation
is more universal,
unrestricted to the case of specific geometry.

\begin{figure}[t]
\begin{tabular}{c}
\includegraphics[width=90mm]{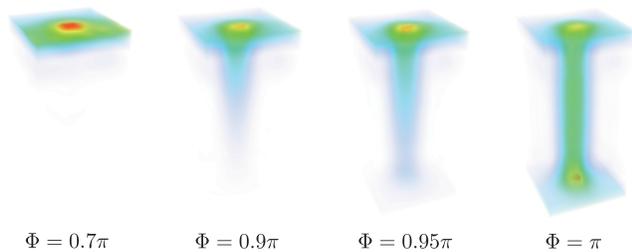}
\end{tabular}
\vspace{-2mm}
\caption{(Color online)
Penetration of the surface wave function along a flux tube of strength $\Phi$.}
\label{invasive}
\end{figure}

There is an inverse effect in 
\textcolor{black}{this specific property} 
of the surface state.
Along a flux tube of strength $\pi$ (half unit flux quantum)
piercing a TI sample
a pair of gapless helical modes bound to the tube is induced. 
These 1D helical channels are shown to be perfectly conducting
\cite{pcc},
and topologically protected as well
\cite{Ran_nphys, TeoKane, disloc}.
In the presence of a surface at which the flux tube is terminated, 
how are these 1D channels connected to the 2D helical surface states?
In Fig. \ref{invasive}
we demonstrate that the noninvasive surface state 
becomes gradually invasive into the bulk 
with the aid of the flux tube.
When the total amount of the flux is not precisely $\pi$,
penetration of the surface state into the bulk is exponentially suppressed.
When the flux is exactly $\pi$, the surface state can penetrate 
into the bulk as deeply as the system's configuration allows it. 
In a sense the $\pi$-flux drags the surface state into the bulk, 
making it {\it invasive}. 

\begin{figure}[t]
\begin{tabular}{c}
\includegraphics[width=70mm]{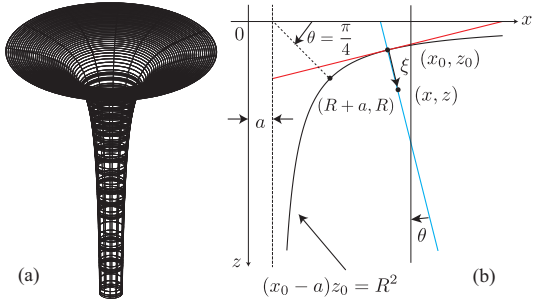}
\end{tabular}
\vspace{-2mm}
\caption{(Color online)
(a) Image of the ``drain''.
(b) Cross section of the hyperbolic surface on the $xz$-plane
(only the $x>0$ part is shown).}
\label{drain}
\end{figure}

This Communication is intended to reveal the nature of this noninvasive metallic state
that appears on topological insulator surfaces
\textcolor{black}{
by demonstrating that the surface state is by itself topologically protected.
This provides with a scenario
alternative to the standard bulk-boundary correspondence picture
that 
attributes 
the same protection to (the non-trivial value of) a bulk topological invariant.}
We start by simulating the behavior of the surface wave function along a flux tube.
Then, as a complementary to this,
we analytically establish the correspondence between the bulk and the surface descriptions.
This is achieved in the second half of the paper,
by employing a configuration in which the surface state 
can partly penetrate into the bulk.
To ease analytic treatments the surface 
is designed to shape a smooth hyperbolic form,
which may look like a ``drain''
[see Fig. \ref{drain}, panel (a)].
Mathematically, this is the locus of a hyperbola 
depicted in Fig. \ref{drain} (b) 
when it revolves around the $z$-axis.
In the limit of sharply edged hole ($R\rightarrow 0$)
this reproduces the situation
described by the tight-binding model
employed in the first part for numerical simulations.

Let us briefly describe the model employed in the 
tight-binding simulation.
The model is based on the following 
3D Wilson-Dirac type
effective Hamiltonian in the bulk
\cite{Liu_nphys, Liu_PRB},
\begin{equation}
H_{\rm bulk} = m(\bm p)\tau_z +
A(p_x \sigma_x + p_y \sigma_y+ p_z \sigma_z)\tau_x,
\label{H_bulk}
\end{equation}
where
$m(\bm p) = m_0 + m_2 \bm p^2$ are Einstein and Newtonian
mass terms encoding a band inversion
due to strong spin-orbit coupling. 
Note that
two types of Pauli matrices $\bm \sigma$ and $\bm \tau$
represent physically real and orbital spins.
It is then implemented on a cubic lattice 
with nearest-neighbor hopping terms.
Periodic boundary conditions are applied
in the $x$- and $y$-directions 
(no surfaces on the corresponding sides),
while the model is restricted in the $z$-direction to
$0 \le z \le N_z-1$.
We consider a system of $N_x\times N_y\times N_z$
and introduce a pair of flux tubes 
piercing RPSRs respectively,
in the $z$ and $-z$-directions at 
$(x,y)=\left(\frac{N_x}{2} - \frac{1}{2}, \frac{N_y}{4} - \frac{1}{2}\right)$ 
and at 
$(x,y)=\left({N_x \over 2} - {1\over 2}, {3N_y\over 4} - {1\over 2}\right)$.
The actual simulation is done in a system of size,
$(N_x, N_y, N_z) = (10, 20, 20)$,
in which
a moderate strength of potential disorder is also included \cite{DWTI}. 
Both the 2D surface and 1D helical modes are shown to be
robust against disorder.

Depicted in Fig. \ref{invasive} is the
evolution of the profile of the lowest energy surface wave functions
when a magnetic flux of different strength $\Phi$ is introduced.
As the flux approaches $\pi$,
the surface state tends to penetrate into the bulk
along a RPSR
(compare different panels of Fig. \ref{invasive} in which
only a half of the system is shown).
When the flux is null, the RPSR is empty.
Yet, one can still hypothesize an electronic motion bound to it.
But then, its energy levitates because of the spin Berry phase $\pi$;
recall half-odd integral quantization of the orbital angular momentum.
Here, since the circumference of the RPSR is atomically small 
($=4a_0$ with $a_0$ being the lattice constant),
the corresponding energy scale of finite-size quantization
is huge.
Clearly, he is no longer compatible with the gapless
(zero-energy) surface state.
The gapless surface state, in turn, does not penetrate into the bulk
along the RPSR.
As the flux is introduced, this Berry phase $\pi$ is
either partly or completely cancelled
depending on the amount inserted.
Then, 
at least a small portion of 1D state along the flux tube
starts to merge with the gapless surface state.
From the viewpoint of the surface state,
a portion of the wave function is dragged 
into the RPSR
\textcolor{black}{
(the wave function gets also accumulated around
the RPSR)}.
This effect should be compared with the asymptotic behavior of 
the analytic formula.

We have seen so far through numerical simulations 
how the surface state loses its noninvasive character
when a flux tube is inserted 
\textcolor{black}{
piercing plaquettes of the bulk crystalline structure.}
We have seen that when the strength of the flux is
precisely $\pi$,
it becomes completely invasive.
These imply, in turn, that the noninvasiveness of the surface state
stems from the Berry phase $\pi$,
which is in a sense omnipresent.
Penetration of the surface state into any hypothetical 
\textcolor{black}{RPSR}
of the lattice
is banned by the existence of this Berry phase $\pi$.

To reinforce the above argument
we formulate this analytically in the remainder of the paper
by solving a corresponding electronic state on
the hyperbolic surface as depicted in Fig. \ref{drain}.
To find the surface Dirac equation on this curved surface
it is convenient to introduce
a set of curvilinear coordinates
$(\xi, \theta, \phi)$ \cite{kado},
defined in terms of the hyperbolic surface:
$\left(\sqrt{x_0^2+y_0^2}-a\right)z_0 =R^2$;
its cross section in the $xz$-plane is shown in Fig. \ref{drain}.
The original cartesian coordinates are expressed as
$x=r\cos\phi$, 
$y=r\sin\phi$,
$z=\xi \cos\theta + R\sqrt{\tan\theta}$,
where
\begin{equation}
r=r(\xi,\theta)=\xi\sin\theta +a+R\sqrt{\cot\theta}
\end{equation}
is an auxiliary parameter dependent on $\xi$ and $\theta$.
The derivatives are represented by
\begin{equation}
\nabla=\bm e_\xi \partial_\xi - {1\over \eta(\theta)-\xi}\bm e_\theta \partial_\theta
+{1\over r(\xi,\theta)}\bm e_\phi \partial_\phi,
\label{grad}
\end{equation}
where the unit vectors $\bm e_\xi$, $\bm e_\theta$, $\bm e_\phi$
are those of the standard 3D polar (spherical) coordinates
\cite{polar}.
$\eta (\theta)$ represents geometrically the radius of curvature of the hyperbolic curve
at $\bm r_0=(x_0, y_0, z_0)$:
\begin{eqnarray}
\eta (\theta) =\sqrt{|\partial_\theta \bm r_0|^2}
= {R\over 2}{1\over\sqrt{\sin^3\theta \cos^3\theta}}.
\label{eta}
\end{eqnarray}
The subsequent analyses are based on
the complex amplitudes of the surface wave function
at the point $(\xi, \theta, \phi)$,
which is vanishingly small 
when $\xi$ significantly exceeds the penetration depth.
If this is much smaller than $R$,
only the range of $\xi \ll R$  is physically relevant.
In this regime we focus on hereafter
apparent singularities in the expressions of Eq. (\ref{grad})
cause no mathematical difficulty.

With the aid of these new coordinates
we deduce the surface Dirac Hamiltonian on the hyperbolic surface
from the bulk effective theory.
In the standard procedure
\cite{aniso, spherical}
this is done by restricting the space of state vectors $|\psi\rangle$
associated with the bulk Hamiltonian $H_{\rm bulk}$
to a set of surface states,
{\it i.e.},
those states that are localized in the vicinity of the hyperbolic surface.
Any surface solution $|\psi\rangle$ of $H_{\rm bulk}$
can be written as a linear combination of two basis solutions,
$|\pm\rangle ={1\over\sqrt{c(\theta)}}
\left(e^{-\kappa_1\xi}-e^{-\kappa_2\xi}\right) |\pm\rangle\rangle$,
{\it i.e.}, $|\psi\rangle=\psi_+ |+\rangle + \psi_- |-\rangle$,
where 
$\psi_\pm$ are (scalar) functions of $\theta$ and $\phi$.
With an appropriate choice of $\kappa_{1,2}$ and 
$|\pm\rangle\rangle$,
$|\pm\rangle$ can be made indeed (two degenerate)
zero-energy eigenstates of $H_{\rm bulk}$
at the ``Dirac point''.
The $\xi$-dependence of the wave function is determined
such that it vanishes 
on the hyperbolic surface. 
The spinor part of the wave function 
$|\pm\rangle\rangle$
can be chosen as
\begin{eqnarray}
|+\rangle\rangle &=& {1\over\sqrt{2}}
\left[\begin{array}{r}
\cos (\theta/2) \\
e^{i\phi} \sin (\theta/2)
\end{array}\right]
\otimes
\left[\begin{array}{c}
1\\ i
\end{array}\right],
\nonumber \\
|-\rangle\rangle &=&  {1\over\sqrt{2}}
\left[\begin{array}{r}
\sin (\theta/2) \\
-e^{i\phi} \cos (\theta/2)
\end{array}\right]
\otimes
\left[\begin{array}{c}
1\\ -i
\end{array}\right].
\end{eqnarray}
Notice that here we have chosen this {\it single}-valued 
[in contrast to the standard SU(2) spinor]
with respect to $\phi\rightarrow\phi +2\pi$.
Though this is a confusing point of this formulation,
whether the basis is double or single valued is simply a matter of choice
\cite{aniso}.
The $\theta$-dependent normalization constant $c(\theta)$ in $|\pm\rangle$
is defined as
\begin{eqnarray}
c(\theta) &=& 
\int_0^\infty d\xi \
r(\xi, \theta) (\eta (\theta) -\xi) \left(e^{-\kappa_1\xi}-e^{-\kappa_2\xi}\right)^2,
\label{c_theta}
\end{eqnarray}
in which 
$r(\eta -\xi)$ is
a measure of the integral associated with 
the volume integral element $r(\eta-\xi)d\xi d\theta d\phi$.
The same measure appears also in the evaluation of
the matrix elements,
$\langle \pm|H_{\rm bulk}|\pm\rangle$
(see below).

The surface Dirac Hamiltonian $H_{\rm surf}$ is
obtained in the spirit of $\bm k\cdot\bm p$-theory
\cite{aniso, spherical, kado}.
Or, in the language of degenerate perturbation theory
this can be regarded as a secular equation
for the coefficients, $\psi_\pm (\theta,\phi)$;
they are solutions of the Dirac equation,
$H_{\rm surf}\bm \psi =E\bm \psi$ with
$\bm\psi = (\psi_+, \psi_-)^T$.
We find the coefficient matrix $H_{\rm surf}$
by evaluating the matrix elements
$\langle\pm|H_{\rm bulk}|\pm\rangle$
as
\begin{equation}
H_{\rm surf} =
\left[
\begin{array}{cc}
\langle +|H_{\rm bulk}|+\rangle& 
\langle -|H_{\rm bulk}|+\rangle\\
\langle +|H_{\rm bulk}|-\rangle&   
\langle -|H_{\rm bulk}|-\rangle
\end{array}
\right]
= \left[
\begin{array}{cc}
0 & D_- \\
D_+ & 0
\end{array}
\right],
\label{H_surf1}
\end{equation}
where
\begin{equation}
D_\pm=
\pm A_\theta{\partial_\theta}
\pm{\partial_\theta A_\theta\over 2}
+ A_\phi \left(-i\partial_\phi +{1\over 2}\right),
\label{D_pm}
\end{equation}
and \cite{m0}
\begin{eqnarray}
A_\theta &=&
{\langle r \rangle \over \langle r (\eta-\xi) \rangle} 
\left( A+ {\left\langle {r\over \eta-\xi} \right\rangle \over \langle r\rangle}m_2 \right)
\equiv {\langle r \rangle \over \langle r (\eta-\xi) \rangle} \tilde{A}_\theta,
\nonumber \\
A_\phi &=&
{\langle \eta-\xi \rangle \over \langle r (\eta-\xi) \rangle} 
\left( A- {\left\langle {\eta-\xi \over r} \right\rangle \sin\theta \over \langle \eta-\xi \rangle} m_2 \right).
\label{A_theta}
\end{eqnarray}
Notice that in Eq. (\ref{D_pm})
the $\phi$-derivative in $D_\pm$
is shifted by $1/2$, which is nothing but the ``Berry phase'' of amount $\pi$.
Since we have {\it chosen} the spinor part of wave function
{\it single}-valued, the orbital angular momentum $L_z$,
defined as
$\bm \psi (\theta,\phi) = e^{i L_z \phi} \bm Z(\theta)$,
takes {\it formally}
an integral value, $L_z = 0, \pm 1, \pm 2, \cdots$.
But due to the Berry phase $\pi$ 
the {\it physical} angular momentum 
$\tilde{L}_z = L_z +1/2$
becomes {\it half-odd} integral
\cite{aniso, prism}.
In Eqs. (\ref{A_theta})
the $\xi$-average $\langle f\rangle$ of a function $f(\xi)$
is defined in terms of a $\xi$-integral
similar to Eqs. (\ref{c_theta}),
{\it i.e.},
\begin{equation}
\langle f\rangle =
{\int_0^\infty d\xi\ f(\xi) \left(e^{-\kappa_1\xi}-e^{-\kappa_2\xi}\right)^2
\over 
\int_0^\infty d\xi \left(e^{-\kappa_1\xi}-e^{-\kappa_2\xi}\right)^2}.
\end{equation}

The effective ``Dirac theory'' on the hyperbolic surface is
prescribed by Eqs. (\ref{H_surf1}), (\ref{D_pm}) and (\ref{A_theta}).
We now attempt to construct zero energy solutions of this effective model.
To ease physical interpretation of the results
it is useful to modify one of the coordinates
by using, instead a (dimensionless) angle $\theta$,
a linear coordinate $\zeta (\theta)$ having the dimension of length
such that
\begin{equation}
\zeta (\theta)=\int_{\pi/4}^\theta d\theta'
{\langle r(\xi,\theta') (\eta (\theta') -\xi) \rangle \over \langle r (\xi, \theta') \rangle}.
\end{equation}
Notice that 
$(\eta (\theta) - \langle\xi\rangle)d\theta$ is a line integral element associated with the locus
of the point $\bm r_0=(x_0, y_0, z_0)$
along a hyperbola at fixed $\phi$.
Thus, at a large distance $r \gg R$ on the surface ($xy$-plane)
from the $z$-axis ($\theta \ll \pi/4$), 
$-\zeta (\theta)$ can be identified as the radial component $r$ of the
standard 2D polar coordinates $(r,\phi)$,
while in the opposite limit of $\theta \gg \pi/4$,
$\zeta (\theta)$ can be identified as $z$, the depth into the 
\textcolor{black}{hole}.
Since $d\zeta /d\theta =  \langle r (\eta-\xi) \rangle / \langle r\rangle$,
the off diagonals in $H_{\rm surf}$ 
[see Eq. (\ref{H_surf1})]
becomes in the $(\zeta,\phi)$-basis,
\begin{equation}
D_\pm=\pm\tilde{A}_\theta \partial_\zeta
\pm{\partial_\zeta \tilde{A}_\theta\over 2}
+A_\phi \left(-i\partial_\phi +{1\over 2}\right).
\label{D_pm2}
\end{equation}

How does the wave function penetrate (or not penetrate) into the 
\textcolor{black}{hyperbolic hole?}
What happens to the Berry phase $\pi$
on the surface sufficiently away from the 
\textcolor{black}{hole?}
Answers to these questions are encoded in the explicit form of
$H_{\rm surf}$.
Let us focus on
the zero energy solutions
for comparison with the result of numerical simulations.
There are two of such solutions,
either with spin up or down,
$\bm\psi_{E=0}^{(\pm)} = e^{i L_\pm \phi} Z_\pm (\zeta)\bm e_\pm$,
where $\bm e_+ = (1,0)^T$, $\bm e_+ = (0,1)^T$,
which satisfy, respectively,
$D_\pm \bm\psi_{E=0}^{(\pm)}=0$.
This can be readily solved as
\begin{equation}
Z_\pm (\zeta) = {1\over\sqrt{\tilde{A}_\theta (\zeta)}}
\exp\left[\mp\tilde{L}_\pm \int_0^\zeta d\zeta'
{A_\phi (\zeta') \over \tilde{A}_\theta (\zeta')}\right],
\label{Z_pm}
\end{equation}
where
$\tilde{L}_\pm = L_\pm+1/2$
\cite{accumulation}.
In the asymptotic limit $\zeta\rightarrow \infty$,
$A_\phi / \tilde{A}_\theta$ in the exponent can be readily approximated as
${A_\phi\over \tilde{A}_\theta} \simeq {1\over \langle r\rangle}\left(1-\left\langle {1\over r}\right\rangle {m_2 \over A}\right)$.
Deep inside the 
\textcolor{black}{hyperbolic hole,}
$\zeta\simeq z$,
and also,
$\langle r\rangle \simeq a + \langle \xi\rangle$ and
$\left\langle {1\over r}\right\rangle \simeq \left\langle {1\over a+\xi}\right\rangle$
become constant,
therefore
Eqs. (\ref{Z_pm})
decay {\it exponentially} 
under the convergence conditions:
$\tilde{L}_+ \geq 1/2$ for $Z_+ (\zeta)$ and 
$\tilde{L}_- \leq -1/2$ for $Z_- (\zeta)$
\cite{half-odd}.
In this regime, the wave function decays
exponentially as it penetrates deeper into the 
\textcolor{black}{hyperbolic hole,}
in other words,
it actually barely penetrates the bulk (noninvasiveness).

How about the opposite limit,
{\it i.e.},
on the surface as $\zeta\rightarrow -\infty$?
In this limit
the profile of the wave functions can be directly
compared with those of the 2D Dirac equation
solved in terms of the Bessel functions $J_n (|E| r /A)$
with the use of the polar coordinates $(r,\phi)$.
And also, we expect that the Berry phase $\pi$ becomes {\it ineffective}
on the surface, which seems {\it a priori} contradictory to
Eqs. (\ref{D_pm}) and (\ref{D_pm2}).
A clue to resolve this discrepancy is
in the normalization of the wave function.
On the 2D surface, the wave function $\bm \psi_{2D} (r,\phi)$
should be normalized in terms of the surface integral element,
$r dr d\phi$,
while in the normalization of $\bm \psi (\zeta, \phi)$ 
this measure $r$ is not taken into account.
Indeed, what should be interpreted as the 2D surface
wave function is
$\bm \psi_{2D} (\zeta,\phi) = 
{\bm \psi (\zeta, \phi)\over \sqrt{\langle r(\zeta) \rangle}}$.
Here, the corresponding effective ``2D Hamiltonian'' $H_{2D}$  for $\bm \psi_{2D}$ is
deduced from $H_{\rm surf}$ by the replacement,
$D_\pm \rightarrow
{\cal D}_\pm = D_\pm \pm {\tilde{A}_\theta\over 2} \partial_\zeta\log\langle r\rangle$,
which can be rewritten as
${\cal D}_\pm=\pm\tilde{A}_\theta \partial_\zeta
\pm{\partial_\zeta \tilde{A}_\theta\over 2}
+A_\phi {\cal L}_\pm$,
by noticing $\tilde{A}_\theta\simeq A$ and $A_\phi\simeq -A/\zeta$
in the limit of $\zeta \rightarrow -\infty$,
where
${\cal L}_+=L_+$,
${\cal L}_-=L_- +1$.
The ${1\over\sqrt{\langle r\rangle}}$ factor in the normalization of $\bm \psi_{2D}$
compensates the effects of Berry phase $\pi$.
Thus, as expected,
the Berry phase $\pi$ is 
shown to be {\it ineffective}
on the flat surface away from the 
\textcolor{black}{hyperbolic hole.}
Since in the present limit,
$A_\phi / \tilde{A}_\theta\simeq 1/ \langle r\rangle$ and
$\langle r(\zeta)\rangle \simeq -\zeta$,
the $\zeta'$-integral in the exponent of Eqs. (\ref{Z_pm})
diverges logarithmically, implying 
${Z_\pm (\zeta)\over\sqrt{\langle r\rangle}} 
\propto |\zeta|^{\pm {\cal L}_\pm}$.
These solutions are bounded
only when ${\cal L}_+ \leq 0$ for $Z_+ (\zeta)$,
and ${\cal L}_- \geq 0$ for $Z_- (\zeta)$.
This implies,
combined with the convergence conditions for the opposite asymptotics,
that the zero energy solution is possible only 
when $L_+ =0$  for $Z_+ (\zeta)$, and when
$L_- =-1$  for $Z_- (\zeta)$.
In these two cases
$Z_\pm (\zeta)$ becomes constant, consistently with the fact
only the zeroth order Bessel function $J_0 (|E| r/A)$ is compatible with the
zero energy condition $E=0$.

Let us finally remark how the introduction of a flux tube piercing the 
\textcolor{black}{hyperbolic hole}
modifies the above argument.
In the extreme case of $\Phi=\pi$,
the Aharonov-Bohm flux $\Phi$ and
the Berry phase $\pi$ (the shift of $1/2$) in Eqs. (\ref{D_pm}) and (\ref{D_pm2}) 
cancel out each other.
As a result, the {\it bare} angular momentum $L_\pm$ 
appears in the exponent of the zero energy solutions (\ref{Z_pm}).
This modifies the asymptotic condition
deeply inside the 
\textcolor{black}{hyperbolic hole}
(in the limit of $\zeta\rightarrow \infty$) to
$L_+ \geq 0$ and $L_- \leq 0$.
In the opposite limit (on the surface away from the 
\textcolor{black}{hole)}
the two solutions behave asymptotically as
${Z_\pm (\zeta)\over\sqrt{\langle r\rangle}} 
\propto |\zeta|^{\pm{\cal L}_\pm}$,
{\it i.e.}, formally as before, but ${\cal L}_\pm$ now replaced with
${\cal L}_\pm=L_\pm \mp1/2$.
The two solutions are legitimate only when
${\cal L}_+ \leq 0$ and ${\cal L}_- \geq 0$.
The only possible choice of $L_\pm$ compatible with these two
asymptotic conditions is
$L_+=L_-=0$.
This signifies that
the wave function deeply inside the 
\textcolor{black}{hyperbolic hole}
stays constant in contrast to the previous case (exponential decay).
The $\pi$-flux tube transforms the surface state {\it invasive},
penetrating the bulk to attain the opposing surface.
The asymptotic behaviors on the surface are modified accordingly,
reproducing those of the Bessel function $J_{-1/2} (|E| r/A)$
in the limit of $E\rightarrow 0$.

The surface state of a topological insulator is always cited as 
being 
\textcolor{black}{a manifestation of the topological non-triviality of the bulk (bulk-boundary correspondence),
while the exotic nature of the surface state itself was apt to be ignored.
Here, in this Communication we have revealed that the surface state is by itself 
topologically protected.
The proposed scenario makes this point explicit,
providing with a viewpoint alternative to the standard
bulk-boundary correspondence picture.}

\begin{acknowledgments}
KI acknowledges Tomi Ohtsuki for stimulating discussions.
The authors are supported by KAKENHI;
K.I. by the ``Topological Quantum Phenomena'' (No. 23103511), and
Y.T. by a Grant-in-Aid for Scientific Research (C) (No. 24540375).
\end{acknowledgments}


\begin{thebibliography}{99}
\bibitem{half-odd}
Recall that $\tilde{L}_\pm$ are quantized to be half-odd integral values.
\end{thebibliography}
\end{document}